\documentstyle[twocolumn,epsf,prl,aps]{revtex}
\newcommand {\be}{\begin{equation}}
\newcommand {\ee}{\end{equation}}
\newcommand {\bea}{\begin{eqnarray}}
\newcommand {\eea}{\end{eqnarray}}
\newcommand {\Eq}[1]{Eq.~({\ref{#1}})}
\newcommand {\Equation}[1]{Equation~({\ref{#1}})}
\newcommand {\Sample}{$Y_{1-x}Pr_{x}Ba_2Cu_3O_{7-\delta}\;$}
\newcommand {\Sampleone}{$Y_{.47}Pr_{.53}Ba_2Cu_3O_{7-\delta}\;$}
\newcommand {\Sampletwo}{$Y_{.47}Pr_{.53}Ba_2Cu_3O_{7-\delta}\;$}
\begin{document}
%\draft
\title
{Dynamics of Flux Creep in Underdoped Single Crystals of
{\boldmath \Sample}.}

\author{T. Stein, G. A. Levin,  and C. C. Almasan}
\address{Department of Physics, Kent State University, Kent  OH 44242}
\author{D. A. Gajewski and M. B. Maple}
\address{Department of Physics and Institute for Pure and Applied Physical
Sciences, University of California, San Diego, La Jolla, CA 92093
\vspace{.5 cm}
%\date{\today}
{\rm \begin{quote}
%\begin{abstract}
Transport as well as magnetic relaxation properties of the mixed state were studied on strongly
underdoped \Sample crystals.  We observed two correlated phenomena - a coupling transition 
and a transition to quantum creep.  The distribution of transport current below the coupling
transition is highly nonuniform, which facilitates quantum creep. We speculate that in the
mixed state below the coupling transition, where dissipation is nonohmic, the current
distribution may be unstable with respect to self-channeling resulting in the formation of very
thin current-carrying layers.
\end{quote}
}
}
%\end{abstract}

\maketitle

%\pacs{Pacs: 74.60.Ge, 74.50.+r, 74.72.Bk  }

\narrowtext
\section{Introduction}

The majority of experiments devoted to the study of dissipation in the mixed state of cuprate
superconductors were performed on optimally doped (maximum critical temperature $T_c$) single
crystals, thin films, or superlatttices.  Meanwhile, in underdoped systems, the normal state
properties exhibit a number of unusual features which quite possibly hold a key to the
understanding of the nature of the normal and superconducting states in cuprates.  Also, as a
result of the lower $T_c$ and upper critical field $H_{c2}$, the dissipation in underdoped
systems can be measured down to much lower reduced temperatures $T/T_c$ in relatively small
fields of a few tesla. In contrast, in optimally doped single crystals, the  dissipation falls
below typically detectable levels at much higher reduced temperatures. Thus, underdoped cuprates
facilitate the study of a broader range of the magnetic field - temperature $H-T$ phase diagram
of the ``vortex matter'' than optimally doped superconductors.  These facts provide a compelling
reason to  undertake a comprehensive study of the mixed state of strongly underdoped cuprates,
including both transport and magnetic relaxation measurements.

One of the outstanding and extensively debated questions is the nature of the coupling transition
in layered vortex systems.  A sharp coupling transition has been observed in
superconductor/insulator multilayers beginning with the pioneering work of Giaever\cite{I.Giaever},
and later in Refs.~\cite{White,N.Y.Fogel}.  In a magnetic field $H$ applied normally  to the
planes, the positions of 2D vortices (pancakes) on neighboring superconducting layers tend to be
uncorrelated at high temperatures.  This vortex state is called 2D liquid.  As the temperature
decreases, the correlation in the direction of the applied magnetic field strengthens and the
vortices tend to form coherent 3D flexible lines.  For high-$T_c$ superconductors, though, the
existence of the coupling transition is not as obvious.  Several groups using a 6-terminal (flux
transformer) technique have arrived at conflicting conclusions. 

Safar et al.\cite {H.Safar} reported the observation of a transition to a 3D liquid in
$YBa_2Cu_3O_{7-\delta}$ which manifests itself as a convergence of voltages generated by the
motion of vortices on opposite faces of the sample.  However, such a strong manifestation of the
coupling transition appears to be the exception rather than the rule.  Other groups have
observed that voltages generated on opposite faces of $Bi_2Sr_2CaCu_2O_8$
\cite{R.Busch,Safar2,Doyle,Keener} and $YBa_2Cu_3O_{7-\delta}$ \cite {Y.Eltsev} single crystals
diverge with lowering temperature, rather than converge.

A point of view which can reconcile the results of these experiments is that a significant 
increase in the correlation length $L_c$ of the vortices in the c-direction indeed takes place
at a well defined temperature, but $L_c$ may remain smaller than the thickness of the sample.
Thus, voltages generated on opposite faces of the sample may never converge in spite of a
macroscopic correlation length. The underdoped cuprates present an opportunity to test this
idea for reasons presented above. If the coupling transition can be proven in an underdoped
system, it must also occur in optimally doped cuprates since the coupling between the
$CuO_2$ bilayers decreases with underdoping.

A second important question in the physics of vortex matter is the possibility of non-activated,
temperature independent, creep due to quantum rather than classical (over the barrier)
relaxation at low temperatures.  Superconductors represent, perhaps, the only system in which
relaxation due to quantum creep is an experimentally accessible phenomenon. Here, a strongly non-equilibrium
macroscopic metastable state relaxes coherently without thermal activation. In contrast, in the
majority of other macroscopic metastable systems, relaxation proceeds as a sequence of a large
number of uncorrelated microscopic steps, requiring thermal activation over an energy barrier.

Yet, the evidence of a magnetic relaxation rate that does not extrapolate to zero as
$T\rightarrow 0$ has not reached a point where experimental data can form a cohesive picture of
the phenomenon.  While non-vanishing magnetic relaxation has been observed in both single
crystals and thin films \cite{Mota,Mota1,vanDalen,Hoekstra}, non-vanishing resistance has
been observed only in ultrathin films \cite{Y.Liu,Ephron,Chervenak}.  This has contributed
to the assertion that temperature independent resistance in films and non-vanishing
low-temperature magnetic relaxation in single crystals are unrelated phenomena. Since these two
types of measurements involve different ranges of current (small currents in transport and
large currents, close to the critical current $J_c$, in magnetic relaxation), they are open to
alternative interpretations, not related to quantum creep. For example, Gerber and Franse
\cite{A.Gerber} have argued that non-vanishing magnetic relaxation at low temperatures may
result from self-heating, so that the local temperature of the sample is higher than that of
the ambient.

The best way to address these issues is to conduct both types of measurements on the same
system. A signature of a temperature independent creep appearing in both transport and magnetic
relaxation measurements at the same temperature would be convincing proof that this phenomenon
is not an artifact and reflects a fundamental change in the relaxation process.

The strongly underdoped \Sample ($T_c\approx 17-21\;K$) system is an excellent candidate for
this study.  One interesting aspect of this low $T_c$ system is that its normal state, revealed
by the suppression of superconductivity by magnetic field, is a two-dimensional
insulator\cite{Levin}.  The reduced dissipation in the normal core of a vortex (due to a large
normal-state resistivity $\rho_n$) increases the mobility of the vortices\cite{Bardeen} and,
therefore, favors quantum tunneling \cite{Caldeira,Blatter,G.Blatter:Review}. Thus, in such an
insulator-superconductor material we can expect that the transition from thermally activated
to quantum creep takes place at higher temperatures than in conventional superconductors
or in more metallic cuprates with higher $T_c$.  

In addition, as shown below, this insulator-superconductor system provides an example of a
drastic departure from the current theoretical understanding of the quantum creep phenomenon.
Extrapolation of the theoretical  results for dirty superconductors leads to the conclusion
that the zero temperature magnetic relaxation rate scales with the zero temperature normal
state conductivity $\sigma_n$. Contrary to this, as shown below, the magnetic relaxation rate
of \Sampletwo remains finite in spite of $\sigma_n(T)\rightarrow 0$ at $T\rightarrow 0$.

We reported the observation of quantum creep in \Sample crystals in a recent Letter\cite{Stein}.
In this paper, we present more data and an extended analysis of both transport and magnetic relaxation
measurements on two strongly underdoped single crystals of \Sampleone with $T_c\approx 17$ and
$21\;K$, respectively.  The two twinned single crystals were grown by a self-flux technique as
described elsewhere \cite{L.M.Paulius}. Typical dimensions are $0.8\times 0.5 \times 0.015\;mm^3$,
with the c-axis of the crystals  oriented along the smallest dimension. 

By performing transport measurements as a function of applied current $I$ and magnetic field
$H$, we demonstrate the existence of a current independent coupling transition temperature
$T^{*}(H)$ preceding the crossover to quantum creep. The dissipation is ohmic above $T^*$ and
non-ohmic below $T^*$. A picture that arises from these observations is that the sample, at
$T<T^{*}$, is divided into two macroscopic regions: a layer near the primary face (where the
current contacts are located) which carries most of the transport current, and the rest of the
sample which remains mostly undisturbed by the current. Inside these layers the vortices are
coupled, with the correlation length comparable to the thickness of the respective layer.  On the
other hand, these two macroscopic regions are decoupled from each other.   As a result, the ratio
$V_p/V_s$ of the primary to the secondary voltage increases  with decreasing temperature several
orders of magnitude over  its value in the normal state.

The crossover to temperature independent (quantum) creep takes place at $T=T_q(H)<T^*(H)$ and only
in the top (current-carrying) layer; the rest of the sample continues to exhibit thermally
activated creep. This leads us  to the conclusion that the upper layer has a thickness of only a
few unit cells similar to ultrathin films and multilayers, the only other systems in which quantum
creep has been observed in transport.  Outside this region, the vortices have much greater length,
which suppresses quantum tunneling because the probability of tunneling decreases exponentially
with the length of the tunneling segment.  We argue below  that  the formation of one or several
very thin channels that carry a current density much greater than average  may be a result of 
non-ohmic dissipation below $T^*$  which can lead to an instability of the current distribution
with respect to  self-channeling.

In order to confirm that the T-independent dissipation is due to quantum tunneling, we performed
magnetic relaxation measurements on a similar crystal of \Sample. The results show a transition
to T-independent relaxation rate at approximately the same temperature $T_q(H)$ as in transport.
In addition, we were able to  determine the characteristic relaxation time which characterizes the
rate of  relaxation  uninhibited by the lack of thermal energy.  The value of this ``escape
time" $\sim\;1\;s$ indicates that the relaxation of the magnetic moment is governed by the
diffusion of vortices from the bulk to the outer edge of the sample.

\section{Transport measurements}

\begin{figure}
\epsfxsize=\columnwidth \epsfbox{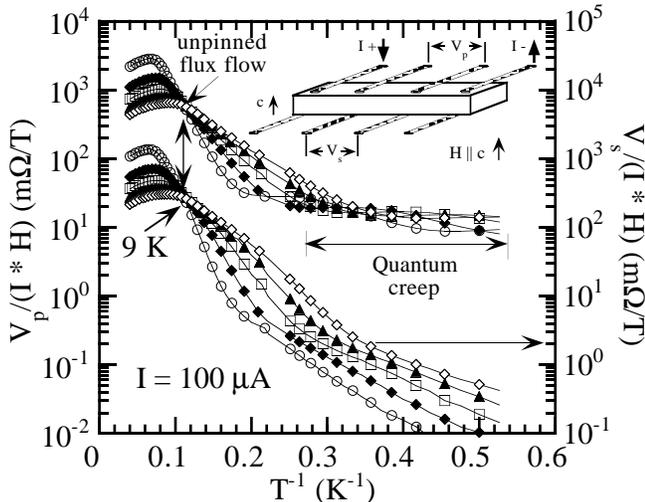}
%\resizebox{\columnwidth}{!}{\includegraphics{99_Stein_PRBfig1.eps}}
\caption{Primary $V_{p}$ and secondary
$V_{s}$ voltages normalized to the total current $I$ and magnetic field $H$ plotted versus
$1/T$  for five magnetic fields ($0.2,\; 0.4,\; 0.6\; 0.8,$ and $1\;T$). The slope decreases
with increasing field which is  parallel to the  $c$-axis.  {\bf Inset}: Contact
configuration used in our measurements.}%
\label{Transport}
\end{figure}

Transport measurements were performed using the ``flux transformer'' contact configuration
(Inset in Fig.~\ref{Transport}).  The current $I$ was injected through the contacts
on one
%
%                Figure #1 {Transport inset Flux_Transformer}
%
\noindent face of the sample and the voltage drops between contacts on the same
(primary voltage $V_p$) and the opposite (secondary voltage $V_s$) faces was
measured for temperature, total current, and magnetic field applied parallel to
the c-axis of the crystal in the ranges of ($1.9\;K\le T\le 20\;K$), ($0.3\;\mu A
\le I\le 2\;mA$), and ($0.2\;T\le H\le 9\;T$) . The single crystal, cleaved into a
bar-shaped sample, was mounted on a single-crystal $MgO$ substrate. ($MgO$
and
$YBa_{2}Cu_{3}O_{7-\delta}$ have similar coefficients of thermal expansion.)  The eight
electrodes were fabricated by bonding 2 mil Au wire to the sample with Ag paste.  A typical
contact resistance was $2\; \Omega$ or less. The mean-field superconducting transition
temperature $T_{c0} \simeq 16.9\;K$ was determined from the temperature dependence of the
in-plane electrical resistance $R(T)$ measured in a magnetically shielded environment ($H <
10^{-2}\;G$) with a low transport current density ($< 10\;A/cm^2$) by using the $2D$ Coulomb
gas model \cite{P.Minnhagen}.  At $T_{c0}$ ($\simeq 16.9\;K$) the resistance of the sample is
$90\%$ that of the normal state value. 

\subsection{Thermally assisted creep}

Figure~\ref{Transport} gives an overall view of the temperature and field dependence of the
primary $V_p$ and secondary $V_s$ voltages.  These voltages are normalized to the current
and field.  The convergence of these curves at $T \approx 9\;K$ indicates a regime where the
dissipation is due to the free motion of vortices.  At lower temperatures, both resistances
exhibit activated $T-$dependence with field-dependent activation energies.  Due to the resistive
anisotropy of the crystal, greater  current flows near the primary face so that $V_{p}>V_{s}$ in
both the normal and mixed states. At even lower $T$, the primary voltage becomes $T$
independent indicating onset of quantum creep while the secondary voltage remains thermally
activated.

At $T\approx 9\;K$, the resistance determined from both the primary and secondary
voltages is proportional to the applied field, i.e.,
\begin{equation}
\label {Bardeen:Stephen}
R_i\propto R_i^{n}\frac{H}{H_{c2}},
\end{equation}
where $R_i\equiv R_{p,s}\equiv V_{p,s}/I$ and $R_i^{n}$ is the corresponding normal state
resistance.  \Equation{Bardeen:Stephen} describes the free flow of vortices  near the upper
critical field $H_{c2}(T)$\cite{Bardeen}.  Fields of $0.2\;T$ and higher are large enough, for
this sample, to shift the onset of free flux flow regime  substantially below the zero field
$T_c\approx 17\; K$. 

\begin{figure}
\
\epsfxsize=.8\columnwidth \epsfbox{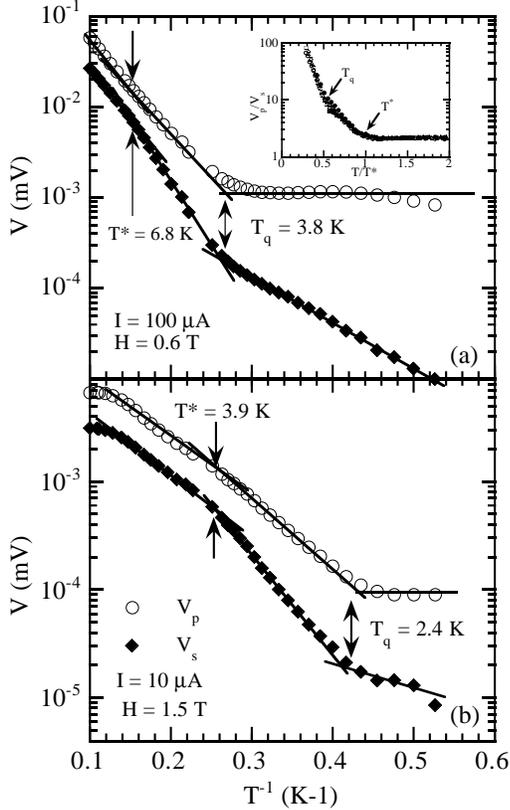}
%\hspace*{.04in}\resizebox{.8\columnwidth}{!}{\includegraphics{99_Stein_PRBfig2.eps}}
\caption{Arrhenius plots of the primary $V_{p}$ and secondary $V_s$ voltages measured in
two different fields and currents;  {\bf (a)} $0.6\;T$ and $I=100\;\mu A$, and {\bf (b)} $1.5\;T$
and $I=10\; \mu A$.  {\bf Inset} to {\bf (a)}: The ``anisotropy" $V_{p}/V_{s}$ plotted
versus reduced temperature $T/T^{*}$ for  applied magnetic fields $H$ of $0.2$, $0.4$,
$0.6$,
$0.8$, $1.0$, $1.5$, $2.0$, $2.5$, and $4.0\;T$.}
\label{Arrhenius}
\end{figure}

The system of vortices undergoes a transition into a new state at a sharply defined temperature
$T^*(H)$ which is the \emph{same} for both primary and secondary voltages and decreases with
increasing $H$ (Figs.~\ref{Arrhenius}(a) and \ref{Arrhenius}(b)).  For $T>T^*$,
the activation energies near the primary and secondary faces of the crystal 
($E_{p,s}\equiv-d\ln V_{p,s}/d(1/T)$) are equal and, therefore, current independent. 
Below $T^*$, both $E_p$ and $E_s$ change [both $V_{p,s}(T)$ curves acquire a different slope], but
always $E_{s}>E_{p}$.  The value of $T^*$ decreases strongly with increasing field.  

It is interesting that the ratio $V_p/V_s$ which changes with temperature and field
appears to scale with the values of $T^*$, as demonstrated in Inset in Fig.~\ref{Arrhenius}(a).
Plotted versus $T/T^*$, the data points for different fields form a single curve. Note also
that the transition to T-independent creep takes place at $T=T_q\approx 0.55-0.6\; T^*$.
Such a strong correlation between $T_q$ and $T^*$ is a clear indication that 
the changes in the vortex system which occur at $T^*$ have a strong impact on the 
transition to quantum creep, or, perhaps, are a prerequisite for such a transition.

\begin{figure}
\epsfxsize=\columnwidth \epsfbox{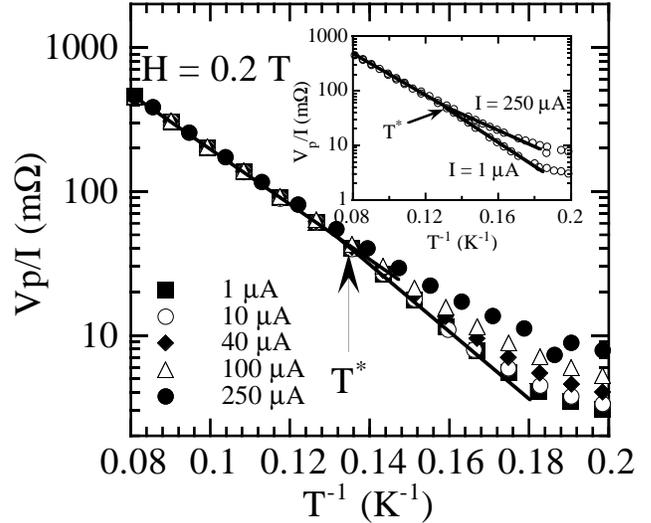}
%\resizebox{\columnwidth}{!}{\includegraphics{99_Stein_PRBfig3.eps}}
\caption{Temperature dependence of the primary resistance $V_p/I$ for different values of
the total current measured in a magnetic field $H=0.2\;T$. The resistance is ohmic above $T^*$
and non-ohmic below $T^*$.  For clarity, the inset shows the same data for two values of
current, $1\mu A$ and $250\mu A$.   The solid lines are guides to the eye.}
\label{Vp(I)}
\end{figure}

Further details of the transformation at $T^*$ are presented in Fig.~\ref{Vp(I)}, where the
primary resistance $R_p(T)$ is shown for several values of the transport current within a
range of over two decades ($1\;\mu A\le I \le 250\;\mu A$). The dissipation at $T>T^*$ is
ohmic, so that the resistance $R_p(T)$ and activation energy $E_p$ are  current independent. 
However, the dissipation becomes non-ohmic below $T^*$.  At low currents, the activation
energy below $T^*$ is {\it greater} than above $T^*$, so that the curve $R_p(1/T)$ has
downward curvature.  The activation energy decreases with increasing current, and, at
sufficiently large currents, $E_{p}(I)$ becomes smaller than it is at $T>T^*$. For large enough
currents, therefore, $R_{p}(1/T)$ acquires upward curvature. This explains the upward
curvature around $T^*$ of $R_p(T)$ in Fig.~\ref{Arrhenius}(a) (large current) and its downward
curvature in Fig.~\ref{Arrhenius}(b) (smaller current). The secondary  voltage $V_s$ reflects 
the substantially smaller current density reaching the secondary (bottom) surface of the
sample. It shows some degree of non-ohmicity, but not as pronounced as the primary voltage.
%
%               Figure  #2 {Arrhenius (a) Low Field inset Anisotropy (b) High Field.}
%
The kink at $T^{*}$ in $R_{p}$ (Figs.~\ref{Arrhenius}(a), \ref{Arrhenius}(b) and \ref{Vp(I)})
is similar to that observed in four-point resistive measurements on $Mo_{77}Ge_{33}/Ge$ and
$Mo/Si$ multilayers \cite{White,N.Y.Fogel} and oxygen deficient $YBa_{2}Cu_{3}O_{7-\delta}$
thin films \cite{X.G.Qiu}. However, the $Mo_{77}Ge_{33}/Ge$ multilayers exhibit a downward
curvature in $R(T)$ \cite{White}, while the $YBa_{2}Cu_{3}O_{7-\delta}$ films show an upward
curvature \cite{X.G.Qiu};  the $Mo/Si$ shows downward curvature for some samples and
upward curvature for others \cite{N.Y.Fogel}.  The data in the Inset in Fig.~\ref{Vp(I)}
demonstrate that the origin of this contradiction is the current dependence of the activation
energy below $T^*$.  The threshold current at which the curvature of $R(T)$ changes sign is
material and sample specific, which explains seemingly contradictory, in this respect, outcomes
of different experiments.

Furthermore, Fig.~\ref{Vp(I)} shows that the crossover temperature $T^*$ is current
independent.  This, along with the fact that $T^*$ is the same for both the primary and
secondary voltages indicates a thermodynamic transition at $T^*$ rather than a kinetic
phenomenon.  All these results indicate that this thermodynamic transition is between 
a system of decoupled 2D vortices and a system of 3D vortices with the macroscopic coherence
length along the direction of the magnetic field (the c-axis in this case).  %
%               Figure  #3 {Vp(I): with inset}
%

\begin{figure}
\epsfxsize=\columnwidth \epsfbox{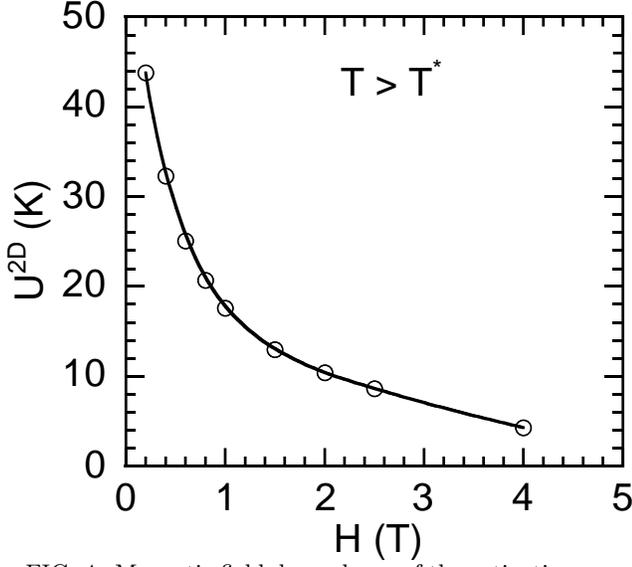}
%\resizebox{\columnwidth}{!}{\includegraphics{99_Stein_PRBfig4.eps}}
\caption{Magnetic field dependence of the activation energy $U^{2D}\equiv d\ln
V_{p,s}/d(1/T)$   determined at  $T>T^{*}$. The solid line is a guide to the eye.}
\label{U2D(H)}
\end{figure}

At $T>T^*$, the vortices are not coherent in the direction of the magnetic field (decoupled) and
behave as 2D ``pancakes''. The dissipation mechanism is activated hopping of 2D pancakes over
potential barriers since the activation energies are the same for the primary and secondary
surfaces in spite of the nonuniform current distribution.  Figure~\ref{U2D(H)} is a plot of the
activation energy $U^{2D}\equiv E_{p,s}$ versus field at $T>T^*$.  $U^{2D}(H)$ decreases
monotonically with increasing field. This is expected because some of the vortices fill the
deepest pinning wells. This leads, due to mutual repulsion, to a smoother potential profile (healing of the random
potential) and, on average, lower activation energies for the rest of the vortices, which provide
the bulk of the dissipation. The decrease of the activation energy due to healing is most
pronounced at low vortex densities (low $H$) since the vortices heal the deepest parts of the
pinning profile first. The efficiency of this process decreases at higher vortex densities (higher
H).  Correspondingly, the rate of change, $dU^{2D}/dH$, decreases with increasing field.   

The pancake vortices form coherent lines at $T<T^*$ and, hence, the activation energy increases
($V_{p,s}$ curve downward).  However, the vortex lines do not extend through the whole thickness
of the sample, so that $V_p$ remains greater than $V_s$ and the ratio $V_p/V_s$ even increases
below $T^*$ (see Inset in Fig.~\ref{Arrhenius}(a)). Moreover, the activation energy decreases
with increasing current and eventually becomes \emph{smaller} than that for 2D pancake
vortices ($V_{p}$ acquires an upward curvature).  This is clearly inconsistent with the idea that
the correlation length $L_c$ is limited by ``flux cutting" processes\cite{Lopez}. If $L_c$ along
the c-axis is destroyed by the stress due to the driving%
%
%               Figure  #4 {U2D(H)}
%
\noindent force, the activation energy  would
decrease with increasing current but could not become smaller than it is for 2D pancake
vortices.  It is obvious then, that below $T^*$ a new channel of relaxation opens up and
becomes dominant at sufficiently large currents.  

Both the field and current dependences of the activation energy at $T< T^{*}$ are consistent
with a $3D$ plastic creep model based on dislocation mediated motion of vortices, similar with
diffusion of dislocations in atomic solids \cite{J.P.Hirth}, with the activation energy given by
\cite{J.P.Hirth,Y.Abulafia}:
\begin{equation}
\label{Upl(I)}
U_{pl}(I,B)=U_{pl}^{o}(B)(1-(\frac{I}{I_{c}^{pl}})^m),
\end{equation}
where $B$ is the magnetic induction, $U_{pl}^{o}(B)$ is the activation energy at $I=0$ and
$I_{c}^{pl}$ is the critical current corresponding to the plastic motion of the vortices. In the
limit of small currents, the activation energy $U_{pl}^{o}$ for the  motion of a dislocation in
a 3D vortex system can be estimated as the energy needed for the formation of a double kink
over the Peierls barrier
\cite{G.Blatter:Review,Y.Abulafia,V.B.Geshkenbein}; i.e.,
\begin{equation}
\label{Upl}
U_{pl}^{o}(B)\simeq\frac{2a_{o}\epsilon_{o}}{\gamma}=
\frac{\Phi_{o}^{2}}{8\pi^{2}\gamma\lambda^{2}_{ab}}
(\frac{\Phi_{o}}{B})^{\frac{1}{2}},
\end{equation}
where $\epsilon_{o}=(\frac{\Phi_{o}}{4\pi\lambda_{ab}})^{2}$ is the line tension for a vortex
aligned along the c-axis, $\Phi_{o}=2.07\times10^{-7}\;G cm^{2}$ is the fluxoid quantum,
$a_{o}=\sqrt{\frac{\Phi_o}{B}}$ is the  vortex lattice constant,
$\gamma=\sqrt{\frac{\lambda_{c}}{\lambda_{ab}}}$ is the anisotropy parameter, and
$\lambda_{ab,c}$ is the penetration depth associated with screening currents flowing in the ab
plane and c-axis, respectively.

Figure~\ref{U3D(H,I)}(a) is a log-log plot of $U^{3D}_s$ versus $H$, extracted from the
secondary voltage $V_{s}(T,H)$ where the current%
%
%               Figure  #5 {U2D(H)}
%
\begin{figure}
\epsfxsize=.9\columnwidth \epsfbox{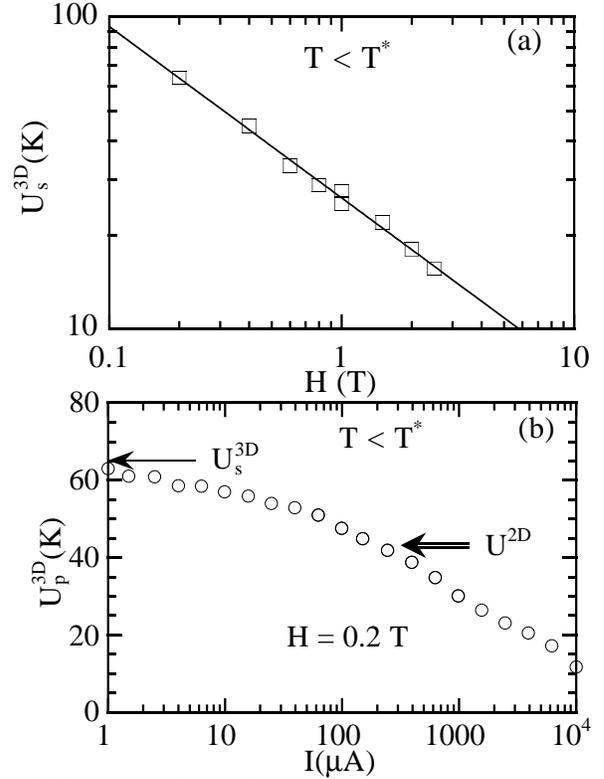}
%\hspace*{.04in}\resizebox{.9\columnwidth}{!}{\includegraphics{99_Stein_PRBfig5.eps}}
\caption{{\bf (a)} Field $H$ dependence of the activation energy $U^{3D}$ determined from 
the secondary voltage as  $d\ln V_{s}/d(1/T)$ for $T<T^*$.  The solid line is a fit of the data to
\Eq{Upl}.{\bf (b)} Current $I$ dependence of the activation energy $U^{3D}_p$ determined from the
primary voltage as  $d\ln V_{p}/d(1/T)$ for $T<T^*$  in a field $H=0.2\;T$. In the limit of small
current, the value of $U^{3D}_p$ is equal to that of $U^{3D}_s$ determined from the secondary
voltage in the same range of temperature $T<T^*$. The value of $U^{2D}$ is also indicated
by the double arrow.  }
\label{U3D(H,I)}
\end{figure}
\noindent is very small ($I\rightarrow 0$). The data exhibit an
$H^{-1/2}$ dependence which is characteristic of the motion of a dislocation in a 3D vortex
structure (\Eq{Upl}).  Figure~\ref{U3D(H,I)}(b) displays the current dependence of the activation
energy $U^{3D}_{p}=E_{p}$ extracted from the primary voltage $V_{p}(T,I)$ for $T<T^{*}$,
measured in a magnetic field $H=0.2\;T$.  The double arrow indicates the value of the current
independent $U^{2D}$ at $T>T^*$ for the same magnetic field.  Notice that $U^{3D}>U^{2D}$ for
$I<0.1\;mA$ and $U^{3D}<U^{2D}$ for $I > 0.1\;mA$. At this threshold current, $V_p(T)$ changes
its curvature from downward open to upward open.  In summary, these results show that the
dissipation at $T<T^{*}$ is determined by two parallel processes: thermally activated  motion  of
correlated vortices (dominant at low currents) with the activation energy greater than that for
a 2D vortex, and  plastic motion of dislocations (dominant at higher currents) with the
activation energy smaller than that for a 2D vortex.

\begin{figure}
\epsfxsize=\columnwidth \epsfbox{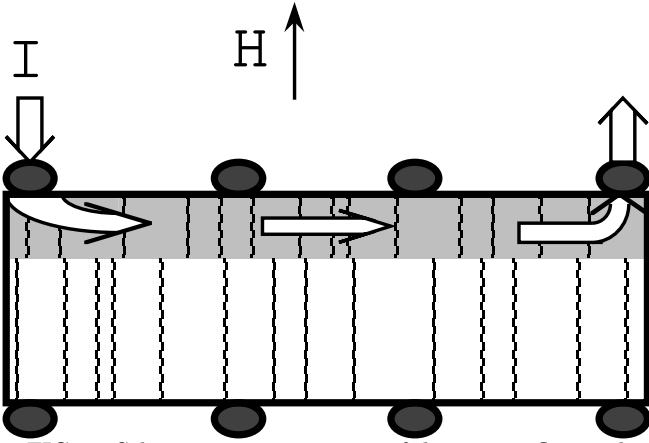}
%\resizebox{\columnwidth}{!}{\includegraphics{99_Stein_PRBfig6.eps}}
\caption{Schematic representation of the current flow within the single crystal. The transport current flows
mainly in the upper (shaded)  layer. The vortices are coupled within each layer, but the  two
layers, current-carrying and ``dormant" (unshaded) are decoupled from each other. The
correlation length of vortices in the dormant layer is greater than that in the current-carrying
layer. }
\label{Model}
\end{figure}

A schematic model of the flux flow which transpires from these observations is shown in
Fig.~\ref{Model}. The current applied through the contacts on the primary face creates a
nonuniform Lorentz force acting on the vortices, 
%
%               Figure  #6 {Model}
%
\noindent which is equivalent to the application of a shear stress to a fragile
solid at $T<T^*$.  The shear stress triggers plastic flow which is strongest near the
primary surface where the current density is greatest.  On the other hand, the vortex lines near the
secondary face remains relatively undisturbed.  This difference of shear stress results in asymetric
flux growth such that the vortices that grow from the secondary surface are longer than
those that originate on the primary face.  These two regions of the sample are uncoupled, yet for
each of them the vortices are coherent over macroscopic distances.  In this scenario, the large
resistive anisotropy measured in the mixed state results from the loss of the phase coherence only
between two macroscopic regions of the sample, not between all microscopic layers (such as $CuO_2$
bilayers) as in the 2D phase at
$T>T^*$.

\subsection{Quantum creep}
At lower temperatures, another transition at a field dependent temperature $T_q$ takes place
(see Figs. 1, 2(a), and 2(b)). The primary voltage $V_{p}$ becomes temperature independent and
scales with the applied magnetic field; i.e., the resistance curves $R_p=V_p/I$ normalized to the
magnetic field $R_p/H$ tend to converge  below $T_q$ (Fig.~\ref{Transport}).  It is important to
note that, although the secondary  voltage (which represents the dissipation in the lower
section of the sample) does not exhibit the transition to T-independent resistance, the
activation energy is noticeably smaller below $T_q$ (Figs.~\ref{Arrhenius}(a) and
\ref{Arrhenius}(b)). 

Since most of the current flows in a thin layer near the primary surface, it is useful to  give an
estimate of the value of the  sheet resistance and residual mobility:
%
%               Figure  #7 {fig:Rbox(I)}
%
\begin{figure}
\epsfxsize=\columnwidth \epsfbox{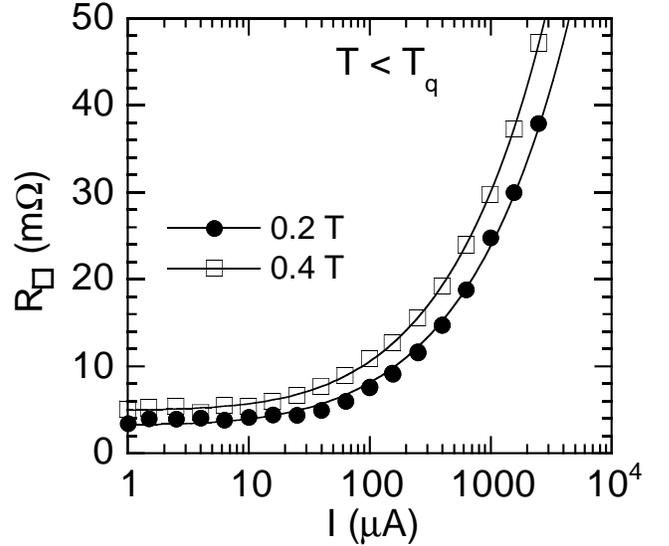}
%\resizebox{\columnwidth}{!}{\includegraphics{99_Stein_PRBfig7.eps}}
\caption{Current $I$ dependence of the sheet resistance $R_{\Box}$ in the quantum creep
regime for two values of magnetic field $0.2$ and $0.4\;T$. }
\label{fig:Rbox(I)}
\end{figure}
\be
\label {Rbox}
R_{\Box}=\frac{R_pb}{\ell }.
\ee
Here $b\approx 0.5 \;mm$ is the  width of the sample and $\ell\approx 0.3\;mm$ is the distance
between the voltage contacts.  Following the conventional treatment \cite{M.Tinkham} of the
dissipation due to vortex drift, the sheet resistance can be expressed in terms of the vortex
mobility $\mu_{res}$ (the total mobility of a moving segment, not the mobility per unit length):
\be
\label {mu}
\mu_{res} =\frac{R_{\Box} c^2}{\phi_0 H},
\ee
where $c$ is the speed of light and $\phi_0$ is the flux quantum.  For $R_{\Box}\propto H$
(Fig.~\ref{Transport}), the vortices below $T_q$  are characterized by a finite, field independent
``residual mobility'' even in the limit $T\rightarrow 0$.  The saturation of $V_{p}$ at low
temperatures persists even at the lowest current of $0.01\;mA$.  At low currents, $R_{\Box}={\cal R}
H$ with ${\cal R} \approx 20 \;m \Omega/ T=6\times 10^{-6} \hbar/e^2\;T^{-1}$; the corresponding
``residual mobility'' $\mu_{res}\approx 1\times10^{10} s/g$. 

As shown in Fig.~\ref{fig:Rbox(I)}, the resistance $R_{\Box}$ is current independent at lower
currents and  increases with increasing current at $I>10\mu A$.  The current
dependence of $R_{\Box}$ can be well fitted with 
\be
\label {Rbox(I)}
R_{\Box}=H{\cal R}\left (1+\frac{I}{I_0}\right )^{1/2},
\ee
with $I_0 \approx 20\mu A$. 

These observations clearly indicate that, in strongly underdoped \Sample, quantum creep
begins to dominate classical, thermally activated creep at relatively  high temperatures
$T_q\sim 5\;K$ in $H=0.2\;T$.  A factor that may facilitate a transition from classical
to quantum creep is the high normal state resistivity.  Previously, we measured the normal
state  resistivity $\rho_n(T)$ of the same sample by suppressing the superconductivity with a
large magnetic field\cite{Levin}.  This showed that the normal state of this superconductor is
insulating, similar to that of $PrBa_{2}Cu_{3}O_{7-\delta}$, so that $\rho_n(T)\rightarrow \infty$
as $T\rightarrow 0$. The reduced dissipation in the normal core of a vortex due to a large
normal-state resistivity increases the mobility of the vortices \cite{Bardeen} and, therefore,
facilitates tunneling\cite{Caldeira}. 

However, current theories cannot be directly applied to this system, because they predict the
Euclidian action $S_E$ of the tunneling process to scale to zero with the zero temperature
normal-state conductivity\cite{G.Blatter}; i.e.,
\be
\label {SE}
\frac{S_E}{\hbar}\approx \frac{\hbar L_c}{e^2\rho_n(0)},
\ee
where $L_c$ is the  length of the tunneling segment.  It is clear, however, from the small
values of the sheet resistance and  residual mobility, $R_{\Box}\propto\mu_{res}\propto
\exp\{-S_E/\hbar\}$, that the Euclidian action does not tend to zero, but remains finite at
$T\rightarrow 0$. Hence, the theory of vortex tunneling, as well as the Bardeen-Stephen
treatment of viscosity must be modified for systems which have insulating normal state
underlying the superconductivity. A serious discrepancy between experimental and theoretical
values of viscosity was also noted in Ref. \cite{Hoekstra} by the analysis of the relaxation
rate in dirty superconductors. 

Previously, the quantum creep in transport measurements was observed exclusively in thin films
whose thickness did not  exceed $30-40\;\AA$ \cite{Y.Liu,Ephron,Chervenak}. The main reason for
this is  the exponential decrease of the probability of tunneling  with increasing correlation
length along the field direction. In films, this length is restricted by the thickness of the
film. Our crystals have a much greater thickness, about $1.5\times 10^5\AA$. However, the fact
that we observe a T-independent primary voltage $V_p$ and a thermally activated $V_s$ indicates
that the thickness of the ``upper" (current-carrying) layer (see Fig.~\ref{Model}) is probably
self-restricted to just a few unit cells, thus facilitating tunneling of such short segments
even at relatively high temperatures ($T\approx 5\;K$). The correlation length is much longer in
the rest of the sample (below the  current-carrying layer),  comparable to the total thickness
of the sample and, as a result, the vortices do not tunnel and  the secondary voltage remains
thermally activated (Figs.~\ref{Transport} and \ref{Arrhenius}).   It is interesting,
however, that  the transition to quantum creep still has an effect on the bulk of the crystal,
because the activation energy determined by the slope  of the secondary resistance $|dlnR_s/d(1/T)|$
decreases at $T_q$. 

\section{Relaxation of magnetization}

To verify that the transition to a temperature independent creep, discussed
above, is due to quantum tunneling, we also performed magnetic relaxation
measurements on another single crystal of \Sampletwo , using a  $SQUID$ magnetometer over
a temperature range $2\;K \leq T \leq 20\;K$ for applied magnetic fields $H$ up
to $5\;T$. A small, $3\; cm$, scanning length was used to minimize the
variations in field strength inside the sample due to spatial inhomogeneities
in the magnet ($\delta H < 0.048\%$).  The superconducting transition
temperature $T_{c} \simeq 21\;K$ of this single crystal was determined from the
onset of diamagnetism measured in a low magnetic field ($H=10\;Oe$). The
irreversibility temperature $T_{irr}$ for a given $H$ was defined as the
temperature above which the zero-field-cooled and field-cooled magnetic moments
are identical. Magnetic  relaxation measurements were performed by cooling the
sample in zero field, applying a field $H+\Delta H$ ($\Delta H=0.3\;T$ for all
$H$) parallel to the c-axis of the crystal and  then reducing it to $H$. The
decay of the resultant  paramagnetic moment was monitored  for several hours (
$\approx 10^4\;s$ ) in  constant field $H$. This procedure was used to  ensure
that the sample was in the fully critical state \cite {Y.Yeshurun}. The
irreversible part of the magnetic moment $M_{irr}$ was obtained approximately by subtracting the
field-cooled moment from the total measured moment.

From this data we can determine whether the relaxation process also exhibits a transition
from thermally activated to quantum relaxation at the same temperature as the transport
resistance.  When the relaxation of the magnetic moment proceeds as a sequence of uncorrelated
microscopic steps, each requiring  thermal activation over an energy barrier, the decay time
$\tau_d$ during which the  induced moment loses a {\it substantial fraction} of its initial
value can be expressed as:
\be
\label {tau(U,T)}
\tau_{d}=\tau_{esc}\exp\left \{ \frac{U(H,T)}{T}\right \}.
\ee
Here the Boltzman factor reflects the degree of availability of energy $U$ required for an
average elementary step to proceed and is essentially independent of the physics of the
relaxation process.  The pre-exponential factor $\tau_{esc}$ is a measure of how rapidly the
relaxation would proceed, had it not been limited by the unavailability of thermal energy. We
call $\tau_{esc}$ an {\it escape time} to distinguish it from the microscopic attempt time
$\tau_a$ which characterizes the period  of vibration of the vortex inside a pinning well.  The
escape time depends on the size of the sample and may depend as well on the magnetic field and 
temperature.  Factorization of the decay time given by \Eq{tau(U,T)} is meaningful as
long as the Boltzman factor $\exp (U/T)\gg 1$, so that it dominates the temperature and field
dependence of $\tau_d$.  Since the activation energy vanishes near $T_c$, the definition of $\tau_d$
can be specified further by taking a linear $T-$dependence of the effective barrier:
\be
\label{U(T)}
U(H,T)\approx U_0(H)\left (1-\frac{T}{T_{cr}}\right ),
\ee
where $T_{cr}$ is the temperature at which the effective activation energy vanishes. It is
commonly taken to be equal to the critical temperature, but may be smaller than $T_c$ and close
to the irreversibility temperature.  Thus, the decay time has the form:
\be
\label {tau(T)}
\tau_{d}=\tau_{esc}\exp\left \{ U_0(H)\left (\frac{1}{T}-\frac{1}{T_{cr}}\right
) \right \}.
\ee

We want to emphasize that \Eq{tau(U,T)} is more general than any
particular  dynamic model of the relaxation process driven by fluctuations.
Therefore, it can also be obtained within the commonly used model  in which the
relaxation is  described as a decay  of the average supercurrent $J$  
determined by current-dependent  activation energy $U(J,B,T)$:
\be
\label {dJ/dt}
\frac{dJ}{dt}=-K\exp\left \{-\frac{U(J,H,T)}{T}\right \}.
\ee
Since  $U(J)$ increases with decreasing current,  this equation can be integrated by
the method of steepest descent:
\be
Kt =\int_{J}^{J_c} dJ^{\prime}\exp\left \{\frac{U(J^{\prime})}{T}\right \}\approx
\frac{T}{|dU/dJ|}\exp\left \{\frac{U(J)}{T}\right \}.
\ee
All unknown parameters can be absorbed into one $\tau_{esc}$ so that 
an approximate solution of \Eq{dJ/dt} has the form
\be
\label {U(T,t)}
\frac{U(J,H,T)}{T}=\ln \left (\frac{t}{\tau_{esc}}\right ),
\ee   
from which \Eq{tau(U,T)} immediately follows.  Equation (13) was obtained earlier\cite{Geshkenbein}
by a more circuitous derivation.

With  $U(J)$ given by the collective creep model\cite{G.Blatter:Review},
\be
\label {U(J):Collective_Creep}
U(J,H,T)=\frac{U(H,T)}{\nu}\left (\left (\frac{J_c}{J}\right )^{\nu} -1\right ),
\ee
\Eq{U(T,t)} gives the following time dependence of $J$:
\be
\label {J(t):Collective_Creep}
J(t)= J_c \left ( 1+ \nu\frac{T}{U}\ln (t/\tau_{esc})\right )^{-1/\nu }.
\ee
On the other hand, for  an arbitrary $U(J)$ in \Eq{U(T,t)}, the initial
decay ($J_c -J \ll J_c$) is linear in logarithm of time (Kim-Anderson formula):
\begin{equation}
\label{Anderson-Kim}
J(t)=J_{c}\left (1-(\frac{T}{U})\ln{\frac{t}{\tau_{esc}}}\right ),
\end{equation}
%
%
%               Figure  #8 {fig:Mirr}
%
\begin{figure}
\epsfxsize=\columnwidth \epsfbox{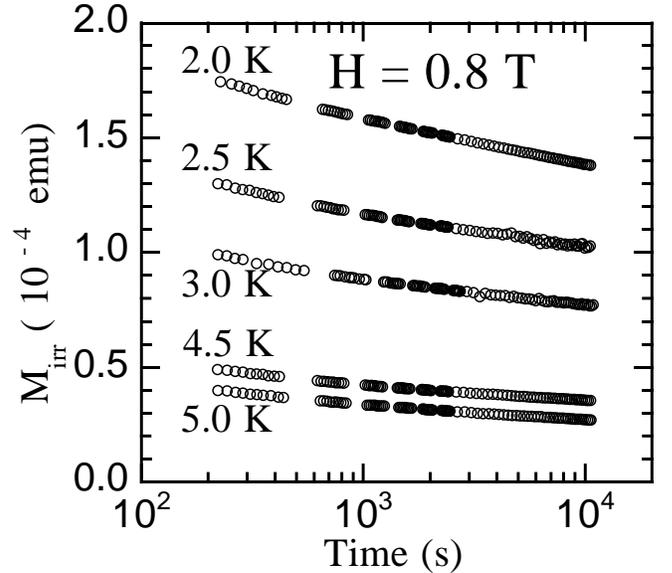}
%\resizebox{\columnwidth}{!}{\includegraphics{99_Stein_PRBfig8.eps}}
\caption{Time $t$ dependence of the relaxation of the irreversible magnetic moment $M_{irr}$ measured at different 
temperatures in a magnetic field $H=0.8\;T$.  To avoid clutter, only a few representative
curves are shown.}
\label{fig:Mirr}
\end{figure}
\noindent where $U=J_c|dU/dJ_c|$, the critical current is determined by the condition
$U(J_c)=0$, and it is assumed that $dU/dJ = const.$ at $J=J_c$. 

Due to the slowness of relaxation, the decay time $\tau_d$ cannot be directly determined by
monitoring the relaxation of the induced moment until it loses a substantial fraction of the
initial value.  An alternative method is to estimate the decay time by extrapolating  the initial 
decay of $J(t)$ to lower  current values.  Specifically, when the initial decay is described 
by  \Eq{Anderson-Kim},  we define $\tau_d$ from the condition  $J(\tau_d)=0$.  
Comparing this definition with
the collective creep formula, Eq.  (15), we see that 
%which leads to \Eq{tau(U,T)}. 
%In the case of collective creep 
%(Eqs.~\ref{U(J):Collective_Creep})~and~\ref{J(t):Collective_Creep}) 
the so defined $\tau_d$ corresponds to a decay to the level of $J_c/(1+\nu)^{1/\nu}$.  
The current density $J$ can be experimentally determined by the irreversible part of the magnetic moment
$M_{irr}(t)\propto J(t)$.

Representative semilog plots of $M_{irr}(t)\propto J(t)$ as a function of time $t$ for several
temperatures measured in a field $H=0.8\;T$ are shown in Fig.~\ref{fig:Mirr}. 
Within a decade of time $10^3-10^4\;s$, the relaxation curves can be well fitted to:
\be
\label {Mirr}
M_{irr}=a-b\ln (t/t_0),
\ee
where $t_0$ is an arbitrary unit of time. The decay time for which $M_{irr}(\tau_d)=0$ is then
given by: 
\be
\label {tau(a,b)}
\tau_d= t_0\exp\left \{\frac{a}{b}\right \}.
%\equiv t_0\exp\left \{\frac{1}{S(t_0)}\right \},
%\tau_d=t_0\exp\{a/b\}\equiv t_0\exp\{1/S(t_0)\},
\ee
With this definition, $\tau_d$ is universal and does not depend on the choice of
$t_0$.   

\begin{figure}
\epsfxsize=\columnwidth \epsfbox{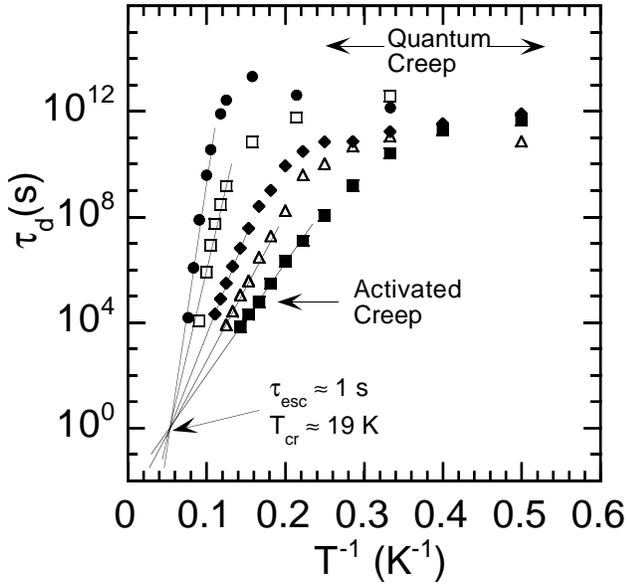}
%\resizebox{\columnwidth}{!}{\includegraphics{99_Stein_PRBfig9.eps}}
\caption{Decay time $\tau_{d}$ versus $1/T$ for several values of magnetic field ($H\;=\; 0.1, \; 0.2, \;
0.6, \; 0.8,$and $1.2\;T$).  The slope decreases with increasing field.  The straight line
extrapolations of the  Arrhenius type dependence converge at $T_{cr}\approx 19\;K$ and 
$\tau_d=\tau_{esc}\approx 1\;s$. The saturation of $\tau_d$ at the level $10^{11}-10^{12}s$ is due
to quantum creep. }
\label{fig:Taud}
\end{figure}

Figure~\ref{fig:Taud} shows the decay time calculated according to \Eq{tau(a,b)} and plotted
against the inverse temperature for
%
%               Figure  #9 {fig:Taud}
%
\noindent different values of magnetic field. A comparison of these
data with
$\tau_d(T)$ given by Eq. (10)  allows us to determine whether the relaxation crosses over
from activated  to non-activated dependence, and at what temperature.  At higher temperatures,
indeed, the data display an Arrhenius dependence with a slope $d\ln \tau_d/d(1/T)$ decreasing
with increasing field.  This trend is consistent with the field dependence of the activation
energy in transport measurements, Fig.~\ref{Transport}. 

It is important to note that the values of the activation energy determined by the slopes $d\ln
\tau_d /d(1/T)$ are not universal and depend on the criterion used to define the decay time (see
\Eq{U(T,t)}). However, the pre-exponential factor $\tau_{esc}$ is universal and can be
determined by the extrapolation of the Arrhenius  dependence of $\tau_d(T)$ to the temperature
$T_{cr}$.  Indeed, a linear  extrapolation of the data in the activated region to
higher temperatures (straight lines in Fig.~\ref{fig:Taud}) shows that  the lines converge at
$T_{cr}\approx 19\;K$ (which is consistent with the value of $T_c$ or $T_{irr}$). The point of
convergence  corresponds to $\tau_d=\tau_{esc}\approx 1\;s$.  This is an extremely large
characteristic time of relaxation in  comparison with the attempt  time which
is typically assumed to be of the order of $10^{-9}-10^{-12}\;s$.  The value of the escape time
can be estimated from the following consideration. The decay of the induced moment occurs when
vortices leave the sample\cite{Feigel'man}. Over long times (in comparison with the microscopic
time scale) any type of rearrangement of vortices reduces to diffusion.  Correspondingly, the
escape time can be estimated as the time required for a vortex to diffuse from the bulk to the
outer edge of the sample:
\be
\label {tauesc}
\tau_{esc}\sim \frac{R^2}{{\cal{D}}_{v}}\sim \frac{R^2}{\omega_a\ell_a^2}\equiv 
\frac{R^2m^{*}}{\hbar},
\ee
where $R$ is the characteristic size of the sample in the direction of
diffusion  (in the $a-b$ plane in our case), and ${\cal{D}}_{v}$ is the
diffusion coefficient determined by the attempt frequency $\omega_a$ and the
average elementary vortex hopping distance $\ell_a$. With $\tau_{esc}\sim 1\;s$
and
$R^2\sim 10^{-2}-10^{-3}\; cm^2$ (for the crystal we measured), \Eq{tauesc} gives
${\cal{D}}_{v}\sim 10^{-2}-10^{-3}\; cm^2/s$. This value of ${\cal{D}}_{v}$ is consistent with
an elementary step of the order of the correlation length $\ell_a\sim 100\;\AA$ and
$\omega_a\sim10^{10}-10^{9}\;s^{-1}$. We define the  effective mass $m^{*}$ of a segment of the vortex
line  through the uncertainty principle, $\omega_a\sim \hbar/m^{*}\ell_a^2$. With these
estimates, the effective mass of the diffusing vortex segment  is $10^2-10^3$ times of the
electron mass. 

\subsection{Quantum creep}
At lower temperatures, the decay time saturates at a roughly temperature and
field-independent level (Fig.~\ref{fig:Taud}). The crossover temperatures $T_q(H)$ from
transport (Fig.~\ref{Transport}) and  magnetic relaxation (Fig.~\ref{fig:Taud}) measurements
are very close in spite of a very large difference in the currents  involved in these
measurements.  The fact that the  transition to a temperature independent
dissipation  takes place in both transport and magnetic relaxation processes, 
and at approximately the same temperature in a given field indicates that both
phenomena have a common origin.

In the regime of  quantum relaxation, the relaxation rate is limited by the probability
of tunneling as determined by the Euclidian action $S_E$. Similar to \Eq{dJ/dt}, the
relaxation rate of the supercurrent can be expressed in terms of the current dependent $S_E(J)$:
\be
\label {dJ/dt(SE)}
\frac{dJ}{dt}=-K\exp\left \{-\frac{S_E(J,H)}{\hbar}\right \},
\ee
which  has a solution similar to \Eq{U(T,t)}:
\be
\label {SE(t)}
\frac{S_E(J)}{\hbar}=\ln \left (\frac{t}{\tau_{esc}}\right ).
\ee
Provided that $dS_E/dJ=const$ at $J_c$, where $J_c$ is determined by the
condition $S_E(J_c)=0$, the initial decay  has same linear in logarithm of time
dependence as in the classical case (see Eq. ~(16)); i.e.,
\be
\label{J(t):Quantum}
J(t)=J_{c}\left (1-\frac{\hbar}{S_0}\ln{\frac{t}{\tau_{esc}}}\right ),
\end{equation}
where $S_0\equiv J_c|dS_E/dJ_c|$. The decay time determined by the extrapolation 
of \Eq{J(t):Quantum} to $J(\tau_d)=0$ is given by:
\be
\label{taud:Quantum}
\tau_d=\tau_{esc}\exp\left \{\frac{S_0}{\hbar}\right\}.
\ee

The value of the escape time should be similar to that in the classical regime since it
is determined by the diffusion uninhibited neither by the lack of 
the thermal energy, nor by the small tunneling probability. Therefore,
we can estimate $S_0$ from the data of Fig.~\ref{fig:Taud}:
\be
\label{So}
\frac{S_0}{\hbar }=\ln \left (\frac{\tau_d}{\tau_{esc}}\right )\approx 25.
\ee
This value is comparable, but somewhat  smaller than those reported for other
systems\cite{Hoekstra}.

\section{Summary and Speculations}
We have observed quantum creep in underdoped \Sample crystals using both transport and magnetic
relaxation measurements. The transition  to quantum creep is preceded by a coupling transition
which leads to non-ohmic dissipation. The  evidence presented in previous sections lead us to a
picture of the current density distribution shown in Fig.~\ref{Model}. Most of the transport
current is confined to a very thin layer below the current contacts.  This current-carrying
layer is decoupled from the rest of the crystal, where the vortices are mostly undisturbed by
the current and are coherent over a macroscopically long distance, perhaps comparable to the
thickness of this ``dormant" layer which is practically the same as the
thickness of the sample. 

The transition from thermally activated to temperature independent dissipation takes place only in
the current-carrying layer.  The relatively large tunneling probability which makes possible the
observation of this crossover at $T \approx 5\;K$ in these crystals is due to very short tunneling
segments, large normal state resistivity, and a large current density.  In the rest of the sample,
the vortices  are much longer and their tunneling is suppressed as manifested by activated
T-dependence of the secondary voltage down to the lowest temperature.  This is consistent 
with the fact that  quantum
creep was previously observed only in transport experiments on ultrathin films and multilayers with
a thickness no more than $30-40\AA$. 

Magnetic relaxation measurements substantiate that this T-independent resistance is due to quantum
creep. The decay time of the magnetic moment becomes T-independent at approximately
the same temperatures (in a given magnetic field) as in transport (see Figs.~\ref{Transport},
\ref{Arrhenius}, and \ref{fig:Taud}). We also determined the characteristic relaxation time
$\tau_{esc}$ which turns out to be very large $\sim 1\;s$ in comparison with the microscopic
attempt time.

While the finding of nonvanishing resistance at $T\rightarrow 0$ in a crystal and its 
correlation with nonvanishing magnetic relaxation rate is important and has never been observed
before, our results raise also another important question.  In the normal and mixed state \emph{
above} the temperature of the coupling transition $T^*$, the crystals of \Sample are not very
anisotropic.  In the sample with the length $L\approx 1\;mm$ and thickness $D\approx 0.015\;mm$,
the ratio $V_p/V_s\sim 2$, (see Inset in Fig.~\ref{Arrhenius}(a)) so that the transport current
fills fairly uniformly the whole cross-section. On the other hand, the transition to quantum
creep indicates that below $T_q$ the current-carrying volume collapses into a thin layer,
possibly just a few unit cells thick.  This favors the  quantum creep for two reasons: minimum
length of the vortex segments and maximum current density which reduces the height of pinning
barriers. This opens the question of the nature  of such a drastic self-channeling
of the transport current. 

A possible answer to this question is related to the non-ohmic, current-dependent resistive
anisotropy. This can be illustrated by the following qualitative dimensional considerations.
According to local electrodynamics, in the sample with thickness $D$ (Inset in
Fig.~\ref{Transport}), the transport current  mostly flows within a layer of thickness
$D_{eff}$\cite{AP}:
\be
\label{Deff}
D_{eff}\approx\frac{D}{\eta };\;\;\; 
\eta\equiv\frac{\pi D}{L}\left (\frac{\rho_c}{\rho_{ab}}\right )^{1/2};\;\;\;
\eta > 1.
\ee
If the effective anisotropy increases with increasing current density $j$ then   
the transport current exhibits a tendency to channel itself into an increasingly narrow layer as
follows.  Let us  assume that to the lowest order in $j$,
the  anisotropy can be written as:
\be
\label{eta}
\eta\approx \eta_0\left (1+\frac{j^2}{j_0^2}\right ),
\ee
where
\be
\label{j(I,Deff)}
j=\frac{I}{D_{eff}},
\ee
and $I$ is the total current.
Combining Eqs. (\ref{Deff}) - (\ref{j(I,Deff)}), we obtain the following
equation for the effective thickness of the current distribution:
\be
\label {Deff(d,I)}
D_{eff}=\frac{D}{\eta_0 (1+\frac{I^2}{D_{eff}^2j_0^2})}.
\ee
This equation has the solution
\be
\label{Deff(d,I):final}
D_{eff}=\frac{D}{2\eta_0 }\left [ 1+\sqrt{1-\frac{I^2}{I_{ins}}}\;\right ],
\ee
where $I_{ins}=Dj_0/2\eta_0$. As the total  current $I$ increases, $D_{eff}(I)$ {\it gradually}
decreases until it reaches half of its zero current value $D/\eta_0$ at $I=I_{ins}$.  For
$I>I_{ins}$, \Eq{Deff(d,I)} does not have a real solution except for $D_{eff}=0$ . Thus, $I_{ins}$ is a threshold  of
instability (the corresponding value of $\eta_{ins}=\eta (I_{ins})=2\eta_0$). For $I>I_{ins}$ there
is no stable current distribution with macroscopic thickness.  The current-carrying layer
compresses itself until it is a few unit cells thick, or until the  current density approaches the
critical value.  The existense  of such an instability would have significant implications
to our understanding of the electrical transport  in layered superconductors and their
applications. 
\section{Acknowledgments}
This research was supported at KSU by the National Science Foundation under
Grant Nos. DMR-9601839 and DMR-9801990,
and at UCSD by U.S. Department of Energy
under Grant No.
DE-FG03-86ER-45230.

\end{document}